\documentclass[twocolumn]{emulateapj}
\pdfoutput=1 
\usepackage{amsmath,amstext}
\usepackage[T1]{fontenc}
\usepackage{amssymb}
\input{epsf}
\usepackage{graphicx}

\usepackage{color}

\newcommand{\ra}[1]{\renewcommand{\arraystretch}{#1}} 

\newcommand{\n}{{\mathrm{n}}}
\newcommand{\p}{{\mathrm{p}}}

\def\be{\begin{equation}}
\def\ee{\end{equation}}
\def\beq{\begin{eqnarray}}
\def\eeq{\end{eqnarray}}

\shorttitle{Core and crust contributions in pulsar glitches: constraints from glitches in the Crab.}
\shortauthors{Haskell et al.}

\begin{document}

\title{Core and crust contributions in pulsar glitches: constraints from the slow rise of the largest glitch observed in the Crab pulsar.}
\author{B.~Haskell}
\affiliation{Nicolaus Copernicus Astronomical Center, Polish Academy of Sciences, ul. Bartycka 18, 00-716 Warsaw, Poland}
\author{V.~Khomenko}
\affiliation{Nicolaus Copernicus Astronomical Center, Polish Academy of Sciences, ul. Bartycka 18, 00-716 Warsaw, Poland}
\author{M.~Antonelli}
\affiliation{Nicolaus Copernicus Astronomical Center, Polish Academy of Sciences, ul. Bartycka 18, 00-716 Warsaw, Poland}
\author{D.~Antonopoulou}
\affiliation{Nicolaus Copernicus Astronomical Center, Polish Academy of Sciences, ul. Bartycka 18, 00-716 Warsaw, Poland}

\begin{abstract}

Pulsar glitches are attributed to the sudden re-coupling of very weakly coupled large scale superfluid components in the neutron star interior. This process leads to rapid exchange of angular momentum and an increase in spin frequency.
The transfer of angular momentum is regulated by a dissipative mutual friction, whose strength defines the spin-up timescale of a glitch. Hence, observations of glitch rises can be used to shed light on the dominant microphysical interactions at work in the high density interior of the star. 
We present a simple analytical model, complemented with more detailed numerical simulations, which produces a fast spin-up followed by a more gradual rise. Such features are observed in some large glitches of the Crab pulsar, including the largest recent glitch of 2017. 
We also use observations to constrain the mutual friction coefficient of the glitch-driving region for two possible locations: the inner crust and outer core of the star. We find that the features of Crab glitches require smaller values of the mutual friction coefficient than those needed to explain the much faster Vela spin-ups. This suggests a crustal origin for the former but an outer core contribution for the latter. 
 \end{abstract}


\section{Introduction} 

Neutron Stars (NSs) offer a unique opportunity to study fundamental physics in an extreme environment. In particular, the ratio of thermal to Fermi energy is generally small enough that neutrons are superfluid throughout most of the star.

Superfluidity has profound consequences for the dynamics of the system, as a superfluid is irrotational and can only mimic rigid body rotation by forming an array of quantised vortices. If vortex motion is impeded by `pinning'  (see e.g.~\citealt{alp77}), the superfluid spin-down lags behind that of the normal component. The excess of angular momentum is then released catastrophically, leading to a fast increase in the observed frequency of the star, or a `glitch' \citep{ai75}.

A complete theory of glitches is still elusive \citep[see][for a recent review of glitch models]{hm15}, despite hundreds of such events having now been observed in over one hundred pulsars \citep[e.g.][]{els+11,Yu13}. One of the pressing open questions is the physical location of the glitch-driving superfluid reservoir. Early models suggested that only the crustal superfluid, where vortices pin to nuclear clusters, is involved in glitches; an expectation in accordance with initial observational estimates that approximately a few percent of the moment of inertia of the star is decoupled \citep{Alpar84a, LEL99}. 

This scenario has been, however, recently re-evaluated by several authors \citep{aghe12,c13,hnl13} who have shown that the crustal superfluid alone cannot account for the measured activity of the Vela pulsar in the presence of strong crustal entrainment \citep[e.g.][]{c12}. 
Similar constraints are obtained when considering the largest glitches in pulsars, which generally require the core's superfluid to be, at least partially, contributing. This issue is still unsettled as there are considerable uncertainties on the entrainment parameters, superfluid gaps and the nature of both the pinning force and the superfluid coupling timescales due to mutual friction in the core and crust of NSs.  These aspects introduce large uncertainties when trying to extract parameters related to the equation of state of NSs \citep{nbh15}, or estimate their masses \citep{hea+15,pah+17}. 

Many pulsars, including the Crab, have spin-up size distributions consistent with powerlaws, and exponentially distributed waiting times \citep{mpw08}. 
On the other hand, the Vela and few other pulsars glitch regularly, with similar spin-up sizes each time\footnote{A catalogue of observed glitches can be found at \url{www.jb.man.ac.uk/pulsar/glitches.html}}. Another such pulsar is PSR J0537-6910, for which there is even a strong correlation between size and waiting time to the next event \citep{mmw+06,aek+18}. 
This difference in behaviour led to suggestions that the glitching mechanism may differ in these two classes of pulsars. Crustquakes might act as triggers in pulsars like the Crab \citep{Onur}, while for Vela, vortex accumulation close to the strongest pinning layers \citep[`snowplow' model,][]{snowplow} can explain the observed behaviour \citep{hps12, HPS2}. 

In this letter we attempt to shed light on the region where pinning occurs by studying the dynamics of the glitch rise. Until recently, the only strong observational constraint on a glitch's rise time came from single-pulse observations of the 2000 Vela glitch, which had a $3-\sigma$ detection limit of $\lesssim40$ seconds \citep{Dodson02}. A similar upper limit is inferred for the 2016 Vela glitch \citep{Vela2016}. In both cases, the spin-up remains unresolved. However, \citet{Vela2016} observed emission changes and excursions of the phase on timescales of seconds, likely linked to the glitch evolution. 

Features of the rise were recently revealed by observations of the largest recorded glitch in the Crab pulsar \citep{Crab2017}, for which a fast rise (inferred to be less than ~6 hours long) is followed by a slower component on a timescale of $\sim 2$ days, a trait that had already been noted in another two large Crab glitches \citep{lsp92,wbl01}. 

\begin{table}
\begin{center}
\ra{1.8} 
 \caption{Observational summary of relevant glitches}
 
\label{table0}
\begin{tabular}{ccrccc}
\toprule
\multicolumn{1}{c}{Glitch} & \multicolumn{1}{c}{$\Delta\nu_{\rm G}$} & \multicolumn{1}{c}{Rise limit} &  \multicolumn{1}{c}{$\Delta{\nu}_{\rm d}$} & \multicolumn{1}{c}{$\tau_{\rm d}$} & \multicolumn{1}{c}{Refs.$^a$} \\

\multicolumn{1}{c}{ } & \multicolumn{1}{c}{($10^{-6}$Hz)} & \multicolumn{1}{c} {$\tau_{\rm r}$} &  \multicolumn{1}{c}{($10^{-7}$Hz)} & \multicolumn{1}{c}{(days)} & \\
\hline
Crab 1989 & $1.85$ &  $0.1$ d$\,\,\,$             & $7$         & $0.8$ & (1)\\ 
Crab 1996 &  $0.66$  &  $0.5$ d$^{*}$        & $3.1$   & $0.5$ & (2)\\ 
Crab 2017 & $14$   &  $0.45$ d$\,\,\,$              & $ 11$  &$1.7$ & (3)\\
Vela 2000 & $35$    & $40$ s$\,\,\,$                  & -- & -- & (4)\\ 
Vela 2017 & $16$    & $48$ s$^{\star}$         & -- & -- & (5)\\
\hline
\end{tabular}
\end{center}
{\sc Notes:}  We present Crab glitches for which a delayed spin-up was observed, and the two Vela glitches for which high resolution data have been used to constrain the rise time. $\Delta\nu_{\rm G}$ denotes the unresolved initial spin-up, with an upper limit $\tau_{\rm{r}}$ for its rise timescale, and $\Delta\nu_{\rm d}$ the delayed component, fitted as an exponential rise with timescale $\tau_{\rm d}$.\\
$^*$ \scriptsize{Value inferred from the timing residuals in \citet{wbl01}}.\\
$^{\star}$ \scriptsize{Taking $(t_3-t_0)$ of \citet{Vela2016} as the maximum spin-up duration.}\\
$^a$ \scriptsize{References: (1)  \citet{lsp92}; (2) \citet{wbl01}; (3) \citet{Crab2017};  (4) \cite{Dodson02}; (5) \cite{Vela2016}.} \\
\end{table}

Such a two stage rise is generally not present in most theories \citep[although note that thermal glitch models predict slower rise for younger pulsars, e.g.][]{ll02}, but can be modelled on general grounds by accounting for vortex accumulation in strong pinning regions, which leads to differential rotation and the propagation of vortex fronts. This naturally produces a slower component of the rise after the initial fast step in frequency \citep{Khomenko}.

In the following we apply the model of \citet{Khomenko} to describe the (slow) rise of the 2017 Crab glitch, and the 2016 glitch of the Vela pulsar. We obtain constraints on the extent of the pinning region and crucially, on the local mutual friction parameters. The results suggest that glitches in the Vela pulsar originate in a higher density region in which mutual friction is stronger, thus strengthening the conclusion that the superfluid in the outer core of neutron stars must be involved. 

\section{Model}

Let us consider a two component star, in which we have superfluid neutrons, labelled with `$\n$', and a charge neutral non-superfluid component of protons and electrons, labelled `$\p$'. If we define their rotation rates as $\Omega_\n$ and $\Omega_\p$ respectively (where we assume that $\Omega_\p$ is the observable rate tracked by the pulsed emission), and their difference, the lag, at time $t$ and cylindrical radius $\varpi$, as $\Delta\Omega (t, \varpi)=\Omega_\p(t, \varpi)-\Omega_\n(t, \varpi)$, we can write an evolution equation of the form \citep{AndComer, Khomenko}:

\be
\frac{\partial \Delta\Omega}{\partial t}= \varpi\tilde{\beta}\Delta\Omega\frac{\partial \Delta\Omega}{\partial \varpi}-2{\tilde{\beta}}\frac{\Omega_\n\Delta\Omega}{(1-\varepsilon_\n)} +\frac{\dot{\Omega}_\infty}{(1-\varepsilon_\n)} \label{mainequation}
\ee

where $\varepsilon_\n$ is the neutron entrainment parameter, $\dot{\Omega}_\infty$ is the contribution of the external spin-down torque and $\tilde{\beta}=\gamma\mathcal{B}/\tilde{\epsilon}$, where we have defined $\tilde{\epsilon}={x_p(1-\varepsilon_n-\varepsilon_p)}/{(1-\varepsilon_n)}$. The parameters $x_\p$ and $\varepsilon_\p$ are the proton fraction and entrainment parameter respectively, $\gamma$ the fraction of free vortices and $\mathcal{B}$ the dimensionless mutual friction parameter. For simplicity, and in order to compare our results directly with observational results from which a pre-glitch spin-down model is subtracted, we will hereafter ignore the last term of (\ref{mainequation}). This is justified as a first approximation for the short timescales involved in the rise, but can have an effect on some of the longer timescales discussed in the following, over which changes in dynamical coupling may lead to spin-down rate changes \citep{hps12, VanessaRise}. 


The mutual friction parameter $\mathcal{B}$ determines the coupling timescale between the two components. In the core, electron scattering off vortex cores is thought to give the strongest contribution, with $\mathcal{B}\approx 10^{-4}$ \citep{als84}. In the crust, vortex interactions with the lattice are likely to dominate, but the exact value of $\mathcal{B}$ is highly uncertain. Current estimates range from $\mathcal{B}\approx 10^{-8}$ for phonon excitations, up to $\mathcal{B}\approx 10^{-1}$ for kelvon mutual friction (for a recent review see e.g. \citealt{HasSed}).

A fundamental parameter that governs the system's dynamics is the fraction of free vortices $\gamma$. This parameter depends on local interactions between neighbouring vortices, and to solve (\ref{mainequation}) an evolution equation for $\gamma$ is needed. In \citet{Khomenko} several prescriptions for the advection of $\gamma$ with the superfluid flow are tested, and found to give qualitatively similar results. Before moving on to the application of these prescriptions, it is, however, instructive to consider a simpler model.

\subsection{Analytic model}

We assume that the normal component is rotating rigidly with frequency $\Omega_\p$, and that the superfluid is perfectly pinned ($\gamma=0$) with a locally uniform lag $\Delta\Omega_{M}$ at the moment before the glitch.  Unpinning of all vortices in this region leads to $\gamma=1$ and according to (\ref{mainequation}) the re-coupling of the two components proceeds over a timescale 
\be
\tau_{\rm r}=\frac{1-\varepsilon_\n}{2\Omega_\p\tilde{\beta}} \,\text{.}
\label{tscale} 
\ee
At the edges of the region over which $\Delta\Omega\longrightarrow 0$ strong differential rotation can develop, turning the first term in equation (\ref{mainequation}) much larger than the second term which is responsible for the rapid recoupling in (\ref{tscale}). Physically this corresponds to vortex re-pinning and accumulation at the edges of the unpinning region, which creates a vortex sheet with high vortex density that most likely drives further unpinning \citep{Lilaunpin, snowplow, Haskell16}. In this case we can simply consider the first term in equation (\ref{mainequation}), set $\gamma=1$, and approximate the equation of motion for the lag $\Delta\Omega$  as:
\be
\label{burgers}
\frac{\partial\Delta\Omega}{\partial t}
\,=\,
\varpi_r \tilde{\beta} \Delta\Omega \frac{\partial{\Delta\Omega}}{\partial\varpi} \, . 
 \ee
 
 If we assume a constant cylindrical radius $\varpi_r\approx 10^6$ cm and an initial condition 
$ \Delta\Omega = H(\varpi -\varpi_0) \, \Delta\Omega_M $, the solution of (\ref{burgers}) takes the form of an outwards travelling fan wave:
 \begin{equation}
 \label{sol2}
 \Delta\Omega= 
 \begin{cases}
    \Delta\Omega_M  (\varpi-\varpi_0) / (v_F \, t)  & \text{for} \quad \varpi<\varpi_F \\ 
    
    \Delta\Omega_M  & \text{for} \quad \varpi\geq\varpi_F 
  \end{cases}
\end{equation}
where $\varpi_0 $ is the initial value of the front's position $\varpi_F$,  which subsequently moves at a speed 
\be
v_F=-\varpi_r \tilde{\beta} \Delta\Omega_M \, .
\label{speed}
\ee
Keeping in mind that in a pulsar $\Delta\Omega$ is negative, we find that $\varpi_F$ moves towards larger radii
(for a detailed description see \citealt{Khomenko}).
 
From angular momentum conservation we have that the size of the glitch $\Delta \Omega_o$ is related to the critical lag in the star $\Delta\Omega_M$ via
\be
\label{glitchsize}
\Delta\Omega_o=-\frac{I_s}{I_T}\Delta\Omega_M \, ,
\ee
where $I_s$ is the moment of inertia of the superfluid reservoir and $I_T\sim10^{45}\; \rm{g\,cm^2}$ the moment of inertia of the star that follows the initial spin-up. 
We approximate $I_s$ as the moment of inertia of a cylindrical shell of constant density $\rho$, at a radius $R_r$ and thickness $\Delta r$, $I_s\approx 4\pi \rho R^4_r \Delta r$. 

If we focus on the delayed component of the rise, i.e. take $\Delta\Omega_o=2\pi\Delta\nu_d$ (see Table \ref{table0}) and assume that the associated timescale $\tau_d$ is the time it takes the vortex front to cross $\Delta r$, we can combine equations (\ref{speed}) and (\ref{glitchsize}) to estimate $\mathcal{B}$:
\be
\mathcal{B} \, = \, \tilde{\epsilon}\,
\frac{I_{T} }{4\pi\rho R_r^5}
\, 
\frac{\Delta\Omega_{d}}{\Delta\Omega_M^2 \tau_d} \, ,
\label{MFestimate}
\ee
 
\section{Results}

Let us now use the solution in (\ref{sol2}) to describe the prolonged spin-up observed in the 2017 Crab glitch (the observations are summarized in Table \ref{table0}). 

We consider two models: a crustal pinning one (denoted with `k' hereafter) and an outer core pinning one (denoted with `c' from now on), with $\rho_k=7\times 10^{13}$ g/cm$^3$ and $\rho_c=3 \times 10^{14}\, \rm{g/cm^3}$ respectively. For the inner crust we take $\tilde{\epsilon}\sim 1$ and for the outer core $\tilde{\epsilon}\sim 0.05$, which is in good agreement with the microphysical models of \citet{c12, PF13}. 
For both cases we take $R_r=11$ km and $\Delta\Omega_M\sim - 10^{-3}\,\rm{rad/s}$ \footnote{
This choice of $\Delta\Omega^k_M$ is motivated by the crustal pinning calculations of \citet{Sevesopin}, however, in the case of the outer core the vortex-flux tubes interactions and respective pinning strengths are less constrained by the theory and could potentially sustain larger lags.}.

Firstly, using equation (\ref{glitchsize}) for the initial, unresolved step $\Delta\Omega_0=\Delta\Omega_G$, we find that in the crust, unpinning must be over a region with  $\Delta_k r \approx 680 \,\rm{m}$. This confirms the expectation that most of the inner crust must be involved in large glitches. For core glitches, this is reduced to $\Delta_c r \approx 150 \,\rm{m}$.  In the case of the smaller, gradual, spin-up  $\Delta\Omega_d$, one has $\Delta_k r \approx 54\, \rm{m}$  and $\Delta_c r \approx 12 \,\rm{m}$.

From equation (\ref{MFestimate}), we find $\mathcal{B}_k\approx 3 \times 10^{-5}$ for the crust, while for the core model $\mathcal{B}_c\approx 4 \times 10^{-7}$. 
 From equation (\ref{tscale}), and setting $\varepsilon_\n=-5$ in the crust and $\varepsilon_\n=0$ in the core, we see that both values of $\mathcal{B}$ would give rise times $\tau_{\rm{r}}\approx0.1$ hours for the unresolved jump, consistent with the observational limit. However, a value of $\mathcal{B}_c\approx 4 \times 10^{-7}$ is much lower than expected for the core, especially in the presence of strong vortex-flux tube interactions \citep{LinkCore, SevesoCore, AlparCore1, AlparCore2}, which suggests a crust model is more likely.
 Furthermore, we note that our model predicts a longer timescale for the delayed rise if the amplitude of that component is larger, as is observed. Applying our method to the other Crab glitches in Table \ref{table0}, we obtain $\mathcal{B}_k\approx 2 \times 10^{-5}$ and $\mathcal{B}_c\approx 3 \times 10^{-7}$ for the 1989 glitch, while for the 1996 event $\mathcal{B}_k\approx 2 \times 10^{-5}$ and $\mathcal{B}_c\approx 2 \times 10^{-7}$, in good agreement with the results for the 2017 glitch.

Let us now shift our attention to the 2016 Vela glitch. The first thing to note is that the $\mathcal{B}$ values obtained for the Crab would lead to rise times $\tau_{\rm r}\approx 1$ hour for the slower rotating Vela pulsar. This is inconsistent with the upper observational limits of $\sim 48$ s of the 2016 (and 2000) Vela glitch. Furthermore, the initial fast rise would be followed by a delayed rise on a similar timescale to that of the Crab, contrary to observations.

We will thus assume that the rise, including any delayed component, must be over before $~$48 s. Given the similarity in sizes,  we use the same size ratio for the regions involved in the glitch for the Vela 2016 glitch as that obtained above for the 2017 Crab glitch. Imposing the constraint $\tau_d<48$ s, we find for Vela $\mathcal{B}_k\gtrsim  0.1$ and $\mathcal{B}_c\gtrsim  10^{-3}$. We note that the latter is a reasonable value of the mutual friction parameter $\mathcal{B}$ for the core, consistent with  expectations from electron scattering off vortex cores \citep{als84, sideryMF}, while the former is close to the upper estimates for kelvon mutual friction in the crust \citep{JonesKelvon, EBKelvon}. For these values  the rapid timescale for the rise is $\tau_r\approx 0.5$ s, which is close to the shorter timescales of few seconds reported in \citet{Vela2016} and to the predictions of more accurate Newtonian and general relativistic results \citep{hps12, Sourie17}. The results are summarized in Table (\ref{table1}).

\begin{table}
\ra{1.8} 
\begin{center}
\begin{tabular}{@{}lll@{}}
\toprule
&  Crab &  Vela \\ 
\vspace{-5mm}&&\\
\hline
\vspace{-4.5mm}&&\\
\multicolumn{2}{@{}l}{Analytic Model:}\\
$\mathcal{B}\,$ in the core: & $4\times 10^{-7}$ & $10^{-3}$\\
$\mathcal{B}\,$ in the crust: & $3\times 10^{-5}$ & $0.1$\\
\vspace{-5mm}&&\\
\hline
\vspace{-4.5mm}&&\\
\multicolumn{2}{@{}l}{Non-linear simulations:}\\
$\mathcal{B}\,$ in the core: & $5\times 10^{-7}-5\times 10^{-6}$ & $3\times 10^{-4}-10^{-3} $\\
$\mathcal{B}\,$ in the crust: & $3\times 10^{-4}-10^{-5}$ & $0.01-0.1$\\
\hline
\end{tabular}
\caption{Summary of results for the mutual friction parameter $\mathcal{B}$ from both the analytic model and the numerical simulations (for which we present the range of limiting values for which the delayed rise is present).}\label{table1}
\end{center}
\end{table}
 
\subsection{Results from non-linear simulations} 

\begin{figure*}
\includegraphics[width=1\columnwidth]{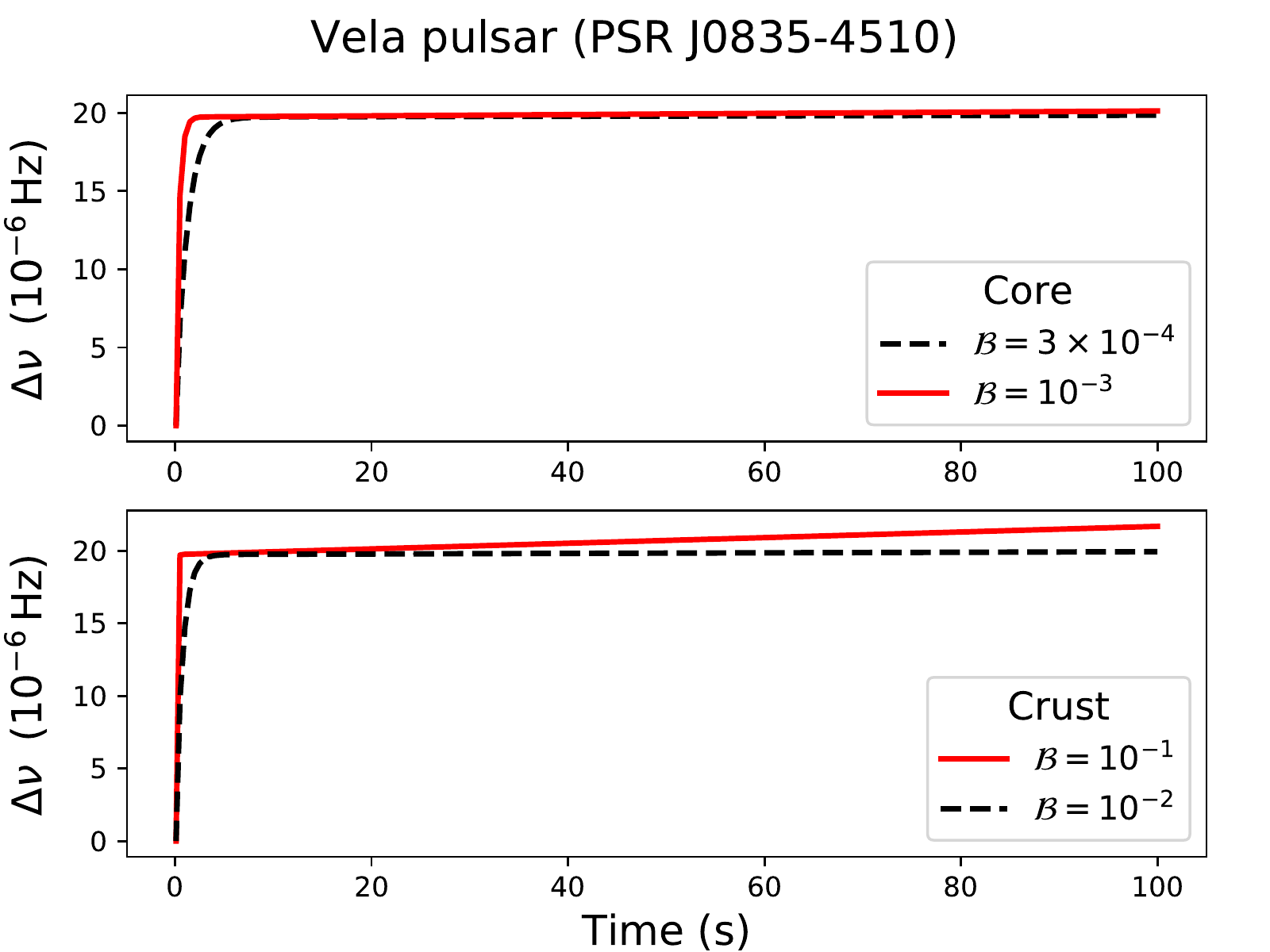}\includegraphics[width=1\columnwidth]{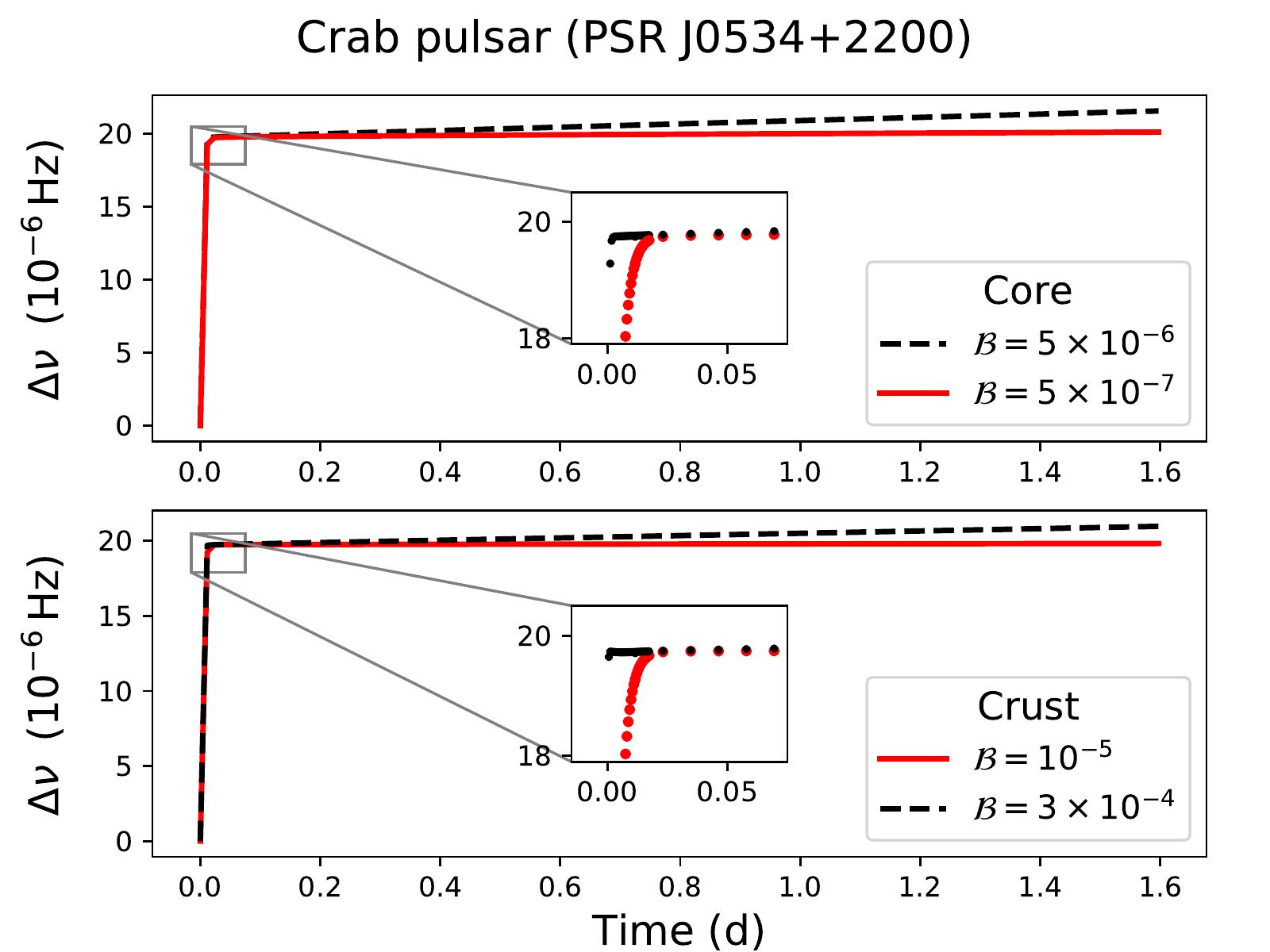}
\caption{\label{fig1} Examples of glitch simulations for the Vela (using $\nu =11 \mbox{ Hz}$) and Crab (using $\nu=30\mbox{ Hz}$). Both models show a rapid rise which, in each pulsar, is followed by a delayed rise for larger values of $\mathcal{B}$. Overall results for the range of $\mathcal{B}$ in which a delayed rise is present are shown in Table \ref{table1}, and confirm that a crustal origin for Crab glitches is in line with theoretical estimates for the mutual friction, while a core contribution is favoured for the Vela pulsar.}

\end{figure*}

We complete the analysis above by running numerical simulations of a full model in which, complementary to (\ref{mainequation}), we provide the following evolution equation for the free vortex fraction $\gamma$:
\be
\frac{\partial \gamma }{\partial t} =\frac{\mathcal{B}}{2 \tilde{\epsilon}}{\varpi}_{0}{\Delta \Omega}_{M}\frac{\partial\gamma}{\partial\varpi} \;\text{.}
\ee

Physically this corresponds to the free vortex density being advected with the unpinning wave.
 
As initial conditions we take $\gamma=1$ in a region $\Delta_r$ with $\gamma=0$ in the rest of the computational domain (of total length L). For numerical convenience we set the ratio $\Delta_r/L=0.67$, which implies that for our simulations the ratio of moments of inertia involved in the glitch is $I_s/I_T\approx\Delta_r/L\approx0.67$. To obtain glitches of a similar size as those observed we thus take an initially constant lag $\Delta\Omega_M = -1.82 \times 10^{-4} \,\rm{rad/s}$. We set $\varpi_0=11$ km and $\tilde{\epsilon}=1$, $\varepsilon_\n=-5$ for crust models, $\tilde{\epsilon}=0.05$, $\varepsilon_\n=0$ for core models.
We evolve the system over the timescales of interest ($\sim 2$ d for the Crab, $\sim 1$ min for the Vela) for a range of $\mathcal{B}$ values, to assess when a delayed rise with amplitude $\Delta\nu_d\approx 0.1\, \Delta\nu_G$ appears. In general, for low values of $\mathcal{B}$ the delayed rise does not develop on the observed timescales, while for high values of $\mathcal{B}$ a large delayed component is visible, as the unpinning wave travels through the simulation region. 


Figure \ref{fig1} presents characteristic examples of the spin evolution. Overall, the results, summarised in Table \ref{table1}, confirm the qualitative picture from the analytic model, suggesting that the crust model is preferred for Crab and the core model for Vela. Note, however, that if the initial pulse in $\gamma$ is set to be $\gamma=1$ over a region $\Delta r$ and $\gamma=0.1$ elsewhere (i.e. imperfect pinning, where $10 \%$ of vortices are free at all time), the linear term in (\ref{mainequation}) always dominates, and no delayed rise is present. If this is the case in Vela but not the Crab, it would explain the lack of a delayed component in the glitches of the former, but is contrary to theoretical expectations that a higher number of free vortices should be present in the Crab, which is younger and hotter than Vela.

\section{Discussion}

In this letter we present a hydrodynamical glitch model that predicts an initial fast rise, followed by a smaller gradual spin-up, as observed in the 2017 Crab glitch \citep{Crab2017} and also two previous glitches \citep{lsp92,wbl01}. 
We then compare our model to observations of Crab and Vela glitches, to estimate the value of the mutual friction parameter ${\mathcal{B}}$, considering two separate cases in which the glitch originates either in the crust or the core of the star.


Our model generally reproduces well the delayed rise observed in several Crab glitches, naturally producing longer timescales for larger glitch amplitudes, and returning consistent values of $\mathcal{B}$ for all glitches. 


We also find that, independently of the model used, the qualitative features of Crab glitches require much smaller values of $\mathcal{B}$. In particular, for core models, we obtain values that are much smaller than what is theoretically expected, while for the crust the results fall in the theoretically predicted range. For the Vela pulsar, on the other hand, larger values are required, which are in line with theoretical estimates for the core but generally at the very upper end of predictions for crustal models.

Theoretically, if we assume that both events originate in the same region of the star it is difficult to justify a larger value of $\mathcal{B}$ in the Vela pulsar. The Vela is older and colder than the Crab pulsar but most mutual friction mechanisms are roughly temperature independent, and even other mechanisms that may be at work, such as coupling of electrons to quasiparticles in vortex cores via interactions the magnetic moment of the neutrons \citep{BE89}, generally become less effective at lower temperatures \citep[see][ for a review of possible mutual friction mechanisms]{HasSed}.

We thus conclude that it is likely that Crab glitches originate in the crust of the star, and Vela glitches in the outer core. This is also supported by the need to consider the core superfluid to explain the observed activity of the Vela pulsar \citep{aghe12,c13,nbh15,hea+15,pah+17}, and would justify the striking difference in the glitch size and waiting time distributions of the two pulsars \citep{mpw08}.

Future high-sensitivity, dense observations of glitches and prompt post-glitch response in other pulsars would allow to confirm whether this interpretation can be validated throughout the pulsar population, and used to explain the general differences between Vela-like pulsars (such as J0537-6910, for which its glitch predictability can be used to plan increased monitoring) and Crab-like pulsars which exhibit a large range in glitch sizes and waiting times.

At the same time, future work on the theoretical front should focus on the vortex-cluster and vortex-flux tube interaction, in order to obtain microphysical constraints on the mutual friction parameters, as are already being developed for pinning forces \citep{Wlazpin}.

\begin{acknowledgments}
We acknowledge support from the Polish National Science Centre grant SONATA BIS 2015/18/E/ST9/00577, and the European Union's Horizon 2020 research and innovation programme under grant agreement No. 702713. Partial support comes from PHAROS, COST Action CA16214.
\end{acknowledgments}
\bibliographystyle{aasjournal}
\bibliography{journals,crab} 

\end{document}